\documentclass[fleqn,twoside]{article}
 \usepackage[headings]{espcrc2}

\usepackage[dvips,final]{graphicx}
 \graphicspath{{./figs-piff/}}

\usepackage{amsmath,amssymb,bm}

\def\muF{\relax\ifmmode\mu_\text{F}^2\else{$\mu_\text{F}^2${ }}\fi}%%%%%%%%%%%%
\def\muR{\relax\ifmmode\mu_\text{R}^2\else{$\mu_\text{R}^2${ }}\fi}%%%%%%%%%%%%
\title{Pion form factor in QCD sum rules, local duality approach,
 and O$(\mathcal A_2)$ fractional analytic perturbation theory}

\author{Alexander~P.~Bakulev%%%
 \address[BLTPh, JINR]{Bogoliubov Laboratory of Theoretical Physics, JINR\\
                       Dubna 141980, Russia\\
                       E-mail: bakulev@theor.jinr.ru}}

\runtitle{Pion FF in QCD SR, LD, and O$(\mathcal A_2)$ FAPT}
\runauthor{Alexander~P.~Bakulev}

\begin{document}

\begin{abstract}
 Using the results on the electromagnetic pion Form Factor (FF)
 obtained in the $O(\alpha_s)$ QCD sum rules with non-local condensates~\cite{BPS09}
 we determine the effective continuum threshold for the local duality approach.
 Then we apply it to construct the $O(\alpha_s^2)$ estimation of the pion FF
 in the framework of the fractional analytic perturbation theory.
\vspace{1pc}
\end{abstract}

\maketitle

In this paper we use the results
on the electromagnetic pion FF
$F_\pi(Q^2)$ in the spacelike region $Q^2=1-10$~GeV$^2$,
obtained in the $O(\alpha_s)$ QCD Sum Rule (SR) approach
with nonlocal condensates (NLC)~\cite{BPS09},
in order to estimate the next-to-next-to-leading (or the $O(\alpha_s^2)$) order
(NNLO) correction
to the pion FF
using both the Fractional Analytic Perturbation Theory (FAPT)
and the Local Duality (LD) approach.
First, we describe the LD approximation in the pion FF calculation
and discuss
how it is possible
to obtain the main ingredient of this approach,
namely, the continuum threshold $s_0(Q^2)$ as a function of $Q^2$.
Then we explain why one need to use FAPT in estimating
the NNLO result for the factorized part of the pion FF.
Next, we describe our model~\cite{BPSS04} for matching function,
gluing soft and hard parts to the complete pion FF.
After that we show how to apply FAPT to estimate the NNLO pion FF
using improved model for matching function.

%%%%%%%%%%%%%%%%%%%%%%%%%%%%%%%%%%%%%%%%%%%%%%%%%%%%%%%%%%%%%%%%%%%%%%%%%%%%%%%
%%%%%%%%%%%%%%%%%%%%%%%%%%%%%%%%%%%%%%%%%%%%%%%%%%%%%%%%%%%%%%%%%%%%%%%%%%%%%%%
\section{Pion FF in the Local Duality approach}
 \label{sec:PiFF.LD}
%%%%%%%%%%%%%%%%%%%%%%%%%%%%%%%%%%%%%%%%%%%%%%%%%%%%%%%%%%%%%%%%%%%%%%%%%%%%%%%
The LD SR~\cite{NR82,Rad95} is produced
from the original QCD SR
in the $M^2\rightarrow \infty$ limit.
For this reason it has no condensate contributions.
The main nonperturbative ingredient in this approach
is the effective continuum threshold $s_0^\text{LD}$ --- it inherits
all the nonperturbative information from the original QCD SR.
At the $(l+1)$-loop order we have
\begin{eqnarray}
 \label{eq:FF.LD}
  F_{\pi}^{\text{LD};(l)}(Q^2,S)
   \equiv
    \int\limits_{0}^{S}\!\!\!\int\limits_{0}^{S}\!\!
     \rho_3^{(l)}(s_1, s_2, Q^2)\,
      \frac{ds_1\,ds_2}{f_{\pi}^2}\,,
\end{eqnarray}
where $S$ should be substituted by the LD effective threshold,
$s_0^{\text{LD};(l)}(Q^2)$,
and $\rho_{3}^{(l)}(s_1, s_2, Q^2)$
is the three-point $(l+1)$-loop spectral density.
In the leading order the integration can be done analytically:
$F_{\pi}^{\text{LD};(0)}(Q^2,S)
  = [S/(4\pi^2f_\pi^2)]\,
    [1 - (Q^2+6S)/(Q^2+4S)\,\sqrt{Q^2/(Q^2 + 4 S)}]$.
The LD prescription for the corresponding correlator \cite{SVZ,Rad95}
implies the relations
\begin{subequations}
\label{eq:LD.s0}
\begin{eqnarray}
\label{eq:LD.s0.LO}
 s_0^{\text{LD};(0)}(0)
  = 4\,\pi^2\,f_{\pi}^2 \simeq 0.7~\text{GeV}^2
\end{eqnarray}
and
\begin{eqnarray}
\label{eq:LD.s0.NLO}
 s_0^{\text{LD};(1)}(0)
  = \frac{4\,\pi^2\,f_{\pi}^2}
         {1+\alpha_s(Q_{0}^2)/\pi}
  \simeq 0.6~\text{GeV}^2\,,~~~
\end{eqnarray}
\end{subequations}
where $Q_{0}^2$ is of the order of $s_0^{\text{LD};(0)}(0)$.
This prescription is a strict consequence of the Ward identity
for the AAV correlator due to the vector-current conservation.
In principle, the $Q^2$ dependence of the LD parameter
$s_0^{\text{LD}}(Q^2)$ (\ref{eq:FF.LD})
should be determined from the QCD SR at $Q^2\gtrsim1$~GeV$^2$.
But as explained in~\cite{BPS09}
the standard QCD SR becomes unstable at $Q^2>3$~GeV$^2$
because of the appearance
of terms in the condensate contributions linearly growing with $Q^2$
\cite{IS83,NR84}.
For this reason, this dependence was known only for
$Q^2\leq 3$~GeV$^2$ and, therefore, most authors usually used the
constant approximation
$s_0^{\text{LD};(0)}(Q^2)\simeq s_0^{\text{LD};(0)}(0)$,
like in~\cite{NR82,BRS00,BPSS04,BO04}, or a slightly $Q^2$-dependent
approximation
$ s_0^{\text{LD};(1)}(Q^2)
\simeq
 4\,\pi^2\,f_{\pi}^2/(1+\alpha_s(Q^2)/\pi)
$,
like in~\cite{BLM07}.

But now, due to the knowledge of the NLC QCD SR prediction~\cite{BPS09}
for the pion FF for $Q^2=1-10$~GeV$^2$,
we can estimate the effective LD threshold
$s_{0}^\text{LD}(Q^2)$
which reproduce these predictions
in the LD approach.
Results are shown in Fig.~1.
It can be represented in this $Q^2$ range
by the following interpolation formula:
\begin{eqnarray}
 \label{eq:FF.LD.s0.App}
  s_{0}^\text{LD}(Q^2)
  = 0.57+0.461\,\tanh\left[\frac{0.0954\,Q^2}{\text{GeV}^2}\right]\,.
\end{eqnarray}
We see that $s_{0}^\text{LD}(Q^2)$
in the mentioned range of $Q^2$
is monotonically increasing function.
Therefore $s_{0}^\text{LD}(Q^2)\neq s_{0}^\text{SR}(Q^2)\approx0.7$~GeV$^2$
and due to this difference the LD approaches of~\cite{BPSS04,BO04,BLM07}
produces significantly lower predictions for $Q^2\,F_\pi(Q^2)$
as compared with QCD SRs with NLC.
\vspace*{-3mm}

%%%%%%%%%%%%%%%%%%%%%%%%%%%%%%%%%%%%%%%%%%%%%%%%%%%%%%%%%%%%%%%%%%%%%%%%%%%%%%%
%%%%   FIGURE 1   %%%%%%%%%%%%%%%%%%%%%%%%%%%%%%%%%%%%%%%%%%%%%%%%%%%%%%%%%%%%%
%%%%%%%%%%%%%%%%%%%%%%%%%%%%%%%%%%%%%%%%%%%%%%%%%%%%%%%%%%%%%%%%%%%%%%%%%%%%%%%
\begin{figure}[htb]
 \centerline{\includegraphics[width=0.47\textwidth]{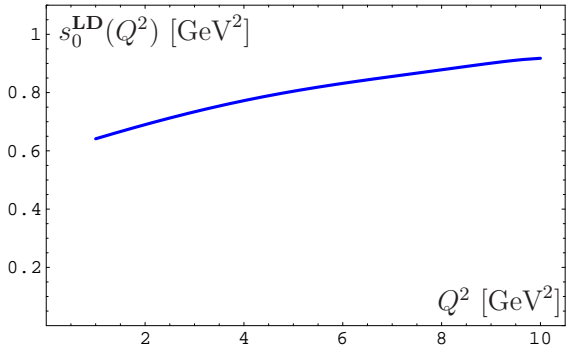}%%%%%%%%%%
  \vspace*{-7mm}}
 \caption{The effective LD threshold $s_{0}^\text{LD}$ as a function
          of $Q^2$. \label{fig:LD.s0}}\vspace*{-7mm}
\end{figure}
%%%%%%%%%%%%%%%%%%%%%%%%%%%%%%%%%%%%%%%%%%%%%%%%%%%%%%%%%%%%%%%%%%%%%%%%%%%%%%%

%%%%%%%%%%%%%%%%%%%%%%%%%%%%%%%%%%%%%%%%%%%%%%%%%%%%%%%%%%%%%%%%%%%%%%%%%%%%%%%
%%%%%%%%%%%%%%%%%%%%%%%%%%%%%%%%%%%%%%%%%%%%%%%%%%%%%%%%%%%%%%%%%%%%%%%%%%%%%%%
\section{Using FAPT for NNLO estimation}
 \label{sec:FAPT}
%%%%%%%%%%%%%%%%%%%%%%%%%%%%%%%%%%%%%%%%%%%%%%%%%%%%%%%%%%%%%%%%%%%%%%%%%%%%%%%
In order to estimate the NNLO contribution to the pion FF
in the QCD SR approach one needs to know the three-loop
spectral density $\rho_{3}^{(3)}(s_1, s_2, Q^2)$
and this is a complicated task.
We want to avoid this tricky calculation
and suggest to use the known collinear two-loop result,
the LD model for the soft part with improved $s_0^\text{LD}(Q^2)$,
and matching procedure of~\cite{BPSS04}.
And one needs to use (F)APT
for the two-loop collinear expression,
as been shown in~\cite{BPSS04,AB08},
in order to have practical independence
with respect to renormalization and factorization scale
setting.

To `glue' together
the LD model for the soft part,
$F_{\pi}^{\text{LD},(0)}(Q^{2})$
(which is dominant at small $Q^2\leq 1$~GeV$^2$),
with the perturbative hard-rescattering part,
$F_{\pi}^{\text{pQCD},(2)}(Q^2)$
(which provides the leading perturbative
 $O(\alpha_s)+O(\alpha_s^2)$ corrections
 and is dominant at large $Q^2\gg1$~GeV$^2$),
in such a way as to ensure the validity of the Ward identity (WI)
$F_{\pi}^{\text{WI};(2)}(0)=1$,
we apply the matching procedure,
introduced in \cite{BPSS04}:
\begin{eqnarray}
  F_{\pi}^{\text{WI};(2)}(Q^{2})
  = F_{\pi}^{\text{LD},(0)}(Q^{2})
 ~~~~~~~~~~~~~~~~~~~~~~ \nonumber\\
 \label{eq:Fpi-Mod.NNLO}
 ~~~ + \left(\frac{Q^2}{2s_0^{(2)}+Q^2}\right)^2
     F_{\pi}^{\text{pQCD},(2)}(Q^2)
\end{eqnarray}
with $s_0^{(2)}\simeq0.6$~GeV$^2$.
In order to test the quality of the matching prescription given by
Eq.\ (\ref{eq:Fpi-Mod.NNLO}), we propose to compare it with the LD
model (\ref{eq:FF.LD}) evaluated at the one-loop order
(i.e., in the $O(\alpha_s)$-approximation~\cite{BO04,BLM07}).
To this end, we construct the analogous $O(\alpha_s)$-model
$F_{\pi}^{\text{WI};(1)}(Q^{2})$,
where we substitute $F_{\pi}^{\text{pQCD},(2)}(Q^2)$
by $F_{\pi}^{\text{pQCD},(1)}(Q^2)=2\,\alpha_s(Q^2)\,s_0^{\text{LD};(0)}(0)/\pi\,Q^2$
and imply the same prescription for the effective LD threshold
as in~\cite{BLM07}, i.~e. (\ref{eq:LD.s0.NLO}).
It is worth to note that the model $F_{\pi}^{\text{WI};(1)}(Q^{2})$,
suggested in~\cite{BPSS04},
works quite well,
although it was proposed without
the knowledge of the exact two-loop spectral density,
which became available later~\cite{BO04}.
The key feature of this matching recipe is that
it uses information on $F_{\pi}(Q^2)$ in two asymptotic regions:
\begin{enumerate}
 \item $Q^2\to0$,
   where the Ward identity dictates $F_{\pi}(0)=1$
   and
   hence $F_{\pi}(Q^2)\simeq F_{\pi}^{\text{LD},(0)}(Q^2)$,
 \item $Q^2\to\infty$,
   where $F_{\pi}(Q^2)\simeq F_{\pi}^{\text{pQCD},(1)}(Q^2)$
\end{enumerate}
in order to join properly the hard tail of the pion FF with
its soft part.
Numerical analysis shows that the applied
prescription yields a pretty accurate result, with a relative error
varying in the range 5\% at $Q^2=1$~GeV$^2$ to 9\% at
$Q^2=3-30$~GeV$^2$.

Now, when the spectral density $\rho_{3}^{(1)}(s_1, s_2, Q^2)$
is known~\cite{BO04},
it is possible to improve the
representation of the LD part by taking
into account the leading $O(\alpha_s)$ correction in the
electromagnetic vertex.
We suggest the following improved WI model
\begin{eqnarray}
  F_{\pi;\text{imp}}^{\text{WI};(1)}(Q^{2},S)
   = F_{\pi}^{\text{LD};(0)}(Q^{2},S)
  ~~~~~~~~~~~~~~~\nonumber\\
  ~~~+ \frac{S}{4\pi^2f_\pi^2}\,
      \frac{\alpha_s(Q^2)}{\pi}\,
       \left(\frac{2S}{2S+Q^2}\right)^2
    \nonumber\\
   + \frac{S}{4\pi^2f_\pi^2}\,
       F^{\text{pQCD},(1)}_\pi(Q^2)\,
       \left(\frac{Q^2}{2S+Q^2}\right)^2
 \label{eq:Fpi-Mod.imp}
\end{eqnarray}
with subsequent substitution $S\to s_0^{\text{LD};(1)}(Q^2)$.
We explicitly display the dependence on the threshold $S$ in
Eq.\ (\ref{eq:Fpi-Mod.imp})---the aim being to apply it later on
with $S=s_0^{\text{LD}}(Q^2)$,
the latter value being extracted by comparing the NLC QCD SR results
with the LD approximation.
Numerical evaluation of this new WI model in comparison with the
exact LD result in the one-loop approximation shows
that the quality of the matching condition is improved:
the relative error is reduced,
reaching 4\% at $Q^2=1-10$~GeV$^2$.

Proceeding along similar lines of reasoning, we construct
the two-loop WI model
$F_{\pi}^{\text{WI};(2)}(Q^{2},s_0^{\text{LD};(2)}(Q^2))$
for the pion FF to obtain
\begin{eqnarray}
  F_{\pi}^{\text{WI};(2)}(Q^{2},S)
   = F_{\pi}^{\text{LD};(0)}(Q^{2},S)
  ~~~~~~~~~~~~~~~\nonumber\\
  ~~~+ \frac{S}{4\pi^2f_\pi^2}\,
        \frac{\alpha_s(Q^2)}{\pi}\,
         \left(\frac{2S}{2S+Q^2}\right)^2
    \nonumber\\
     + \frac{S}{4\pi^2f_\pi^2}\,
        F^{\text{FAPT},(2)}_\pi(Q^2)\,
         \left(\frac{Q^2}{2S+Q^2}\right)^2,
 \label{eq:Fpi.WI-Mod.NNLO}
\end{eqnarray}
where $F^{\text{FAPT},(2)}_\pi(Q^2)$ is the analyticized
expression generated from $F^{\text{pQCD},(2)}_\pi(Q^2)$
using FAPT (see Refs.\ \cite{BMS-APT,BKS05,AB08})
to get a result which appears to be very close
to the outcome of the default scale setting
($\muR=\muF=Q^2$),
investigated in detail in \cite{BPSS04} in the APT approach.
FAPT is needed here in order to obtain analytic expressions
for the pion FF in both possible cases of factorization scale setting:
\begin{enumerate}
 \item[(i)] For $\muF=Q^2$
    there appear factors of the type
    $\left[\alpha_s(Q^2)\right]^{\nu}$
    with fractional powers $\nu=\gamma_n/(2\,b_0)$
    due to the pion distribution amplitude evolution;
 \item[(ii)] For $\muF=\textsl{const}$
    there appears factor
    $\left[\alpha_s(Q^2)\right]^{2}\ln(Q^2/\muF)$.
\end{enumerate}
In any case, the NNLO correction
involves the analytic image of the second power of the coupling,
$\mathcal A_2(Q^2)$.
For this reason we name the whole $F^{\text{FAPT},(2)}_\pi(Q^2)$ term
as $O(\mathcal A_2)$ contribution.\vspace*{-3mm}

%%%%%%%%%%%%%%%%%%%%%%%%%%%%%%%%%%%%%%%%%%%%%%%%%%%%%%%%%%%%%%%%%%%%%%%
%%%  FIGURE 2  %%%%%%%%%%%%%%%%%%%%%%%%%%%%%%%%%%%%%%%%%%%%%%%%%%%%%%%%
%%%%%%%%%%%%%%%%%%%%%%%%%%%%%%%%%%%%%%%%%%%%%%%%%%%%%%%%%%%%%%%%%%%%%%%
\begin{figure}[h]
    \centerline{\includegraphics[width=0.47\textwidth]{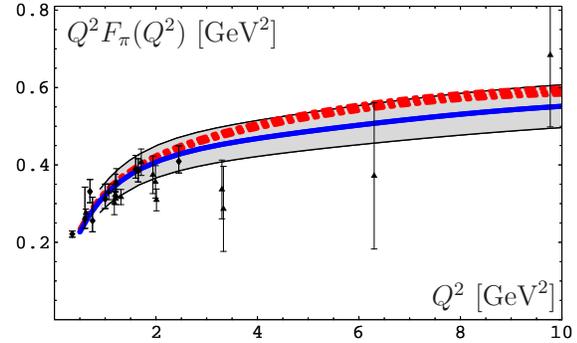}}
  \vspace*{-7mm}
     \caption{\label{fig:FF.LD.2Loop}
     We show as a narrow dash-dotted strip the predictions
     for the pion FF, obtained in the two-loop WI model,
     Eq.\ (\ref{eq:Fpi.WI-Mod.NNLO}), using the improved Gaussian model.
     The width of the strip is due to the variation of the Gegenbauer
     coefficients $a_2$ and $a_4$ (needed to calculate the collinear
     part $F_{\pi}^{\text{pQCD},(2)}(Q^2)$)
     in the corresponding shaded bands for the pion DA
     (indicated by the central solid line).
     Note that this dash-dotted strip shows the effect of
     $O(\mathcal A_2)$ correction
     only for the central solid curve of the shaded band.
}\vspace*{-3mm}
\end{figure}
%%%%%%%%%%%%%%%%%%%%%%%%%%%%%%%%%%%%%%%%%%%%%%%%%%%%%%%%%%%%%%%%%%%%%%%

Interesting to note here,
that in the case of the one-loop approximation
the relative error of WI model (\ref{eq:Fpi-Mod.imp})
appears to be of the order of 10\%.
The relative $O(\alpha_s^2)$-contribution to the pion FF
is of the order of 10\%, as has been shown in~\cite{BPSS04,AB08}.
Hence, the relative error of our estimate is of the order of
1\%---provided we take into account the $O(\alpha_s)$-correction
exactly via the specific choice of $s_0(Q^2)$,
as done in (\ref{eq:FF.LD.s0.App}).

The results obtained for the pion FF with our two-loop
model, i.e.,
Eq.\ (\ref{eq:Fpi.WI-Mod.NNLO}),
and using the effective LD thresholds
$s_{0}^\text{LD}(Q^2)$,
are displayed in Fig.~2.
We see from this figure
that the main effect of the NNLO correction
peaks at $Q^2\gtrsim4$~GeV$^2$, reaching the level of $3-10$\%.

%%%%%%%%%%%%%%%%%%%%%%%%%%%%%%%%%%%%%%%%%%%%%%%%%%%%%%%%%%%%%%%%%%%%%%%%%%%%%%%%%%
%%%%%%%%%%%%%%%%%%%%%%%%%%%%%%%%%%%%%%%%%%%%%%%%%%%%%%%%%%%%%%%%%%%%%%%%%%%%%%%%%%
\section*{Conclusions}
%%%%%%%%%%%%%%%%%%%%%%%%%%%%%%%%%%%%%%%%%%%%%%%%%%%%%%%%%%%%%%%%%%%%%%%%%%%%%%%%%%
\begin{itemize}
  \item  We showed here that the local duality model for pion FF suffers
   from the threshold $s_0(Q^2)$ uncertainty.
   We fixed this uncertainty by demanding that the LD model
   with right setting for $s_0(Q^2)$
   should reproduce results for pion FF obtained in the Borel SRs with NLC~\cite{BPS09}.
   Our results show that $s_0^\textbf{LD}(Q^2)$ grows with $Q^2$.
  \item Our rough model for matching function in~\cite{BPSS04}
   appears to be of good quality ($\approx 10\%$)
   and we managed to improve it here to have quality of $\approx 5\%$.
  \item Using FAPT and improved matching function
   we estimated NNLO correction to the pion FF
   to be of the order of $\approx 3-10\%$.
  \item Our strip of predictions for the pion FF is in a good agreement
   with existing experimental data of Cornell~\cite{FFPI-Cornell} and JLab~\cite{JLab08II}.
\end{itemize}
%%%%%%%%%%%%%%%%%%%%%%%%%%%%%%%%%%%%%%%%%%%%%%%%%%%%%%%%%%%%%%%%%%%%%%%%%%%%%%%%%%

%%%%%%%%%%%%%%%%%%%%%%%%%%%%%%%%%%%%%%%%%%%%%%%%%%%%%%%%%%%%%%%%%%%%%%%%%%%%%%%
%%%%%%%%%%%%%%%%%%%%%%%%%%%%%%%%%%%%%%%%%%%%%%%%%%%%%%%%%%%%%%%%%%%%%%%%%%%%%%%
\section*{Acknowledgments}
This work was supported in part by
the Russian Foundation for Fundamental Research,
grants No.\ ü~07-02-91557, 08-01-00686, and 09-02-01149,
the BRFBR--JINR Cooperation Programme,
contract No.\ F08D-001,
the Deutsche Forschungsgemeinschaft
(Project DFG 436 RUS 113/881/0-1),
and
the Heisenberg--Landau Programme under grant 2009.
%%%%%%%%%%%%%%%%%%%%%%%%%%%%%%%%%%%%%%%%%%%%%%%%%%%%%%%%%%%%%%%%%%%%%%%%%%%%%%%
%%\bibliographystyle{myplb}
%%\bibliography{pion,lambda,qft,nonloc}

\begin{thebibliography}{10}

\bibitem{BPS09}
 A.~P. Bakulev, A.~V. Pimikov, and N.~G. Stefanis,
  Phys. Rev. D79  (2009)  093010.
   %%CITATION = 0904.2304;%%.

\bibitem{BPSS04}
 A.~P. Bakulev, K. Passek-Kumeri\v{c}ki, W. Schroers, and N.~G. Stefanis,
  Phys. Rev. D70  (2004)  033014, 079906(E).
   %%CITATION = HEP-PH 0405062;%%.

\bibitem{NR82}
 V.~A. Nesterenko and A.~V. Radyushkin,
  Phys. Lett. B115  (1982)  410.
   %%CITATION = PHLTA,B115,410;%%.

\bibitem{Rad95}
 A.~V. Radyushkin,
  Acta Phys. Polon. B26  (1995)  2067.
   %%CITATION = HEP-PH 9511272;%%.

\bibitem{SVZ}
 M.~A. Shifman, A.~I. Vainshtein, and V.~I. Zakharov,
  Nucl. Phys. B147  (1979)  385;
   %%CITATION = NUPHA,B147,385;%%.
  \textit{ibid.} 448;
   %%CITATION = NUPHA,B147,448;%%.
  \textit{ibid.} 519.
   %%CITATION = NUPHA,B147,519;%%.

\bibitem{IS83}
 B.~L. Ioffe and A.~V. Smilga,
  Nucl. Phys. B216  (1983)  373.
   %%CITATION = NUPHA,B216,373;%%.

\bibitem{NR84}
 V.~A. Nesterenko and A.~V. Radyushkin,
  JETP Lett. 39  (1984)  707.
   %%CITATION = JTPLA,39,707;%%.

\bibitem{BRS00}
 A.~P. Bakulev, A.~V. Radyushkin, and N.~G. Stefanis,
  Phys. Rev. D62  (2000)  113001.
   %%CITATION = HEP-PH 0005085;%%.

\bibitem{BO04}
 V.~V. Braguta and A.~I. Onishchenko,
  Phys. Lett. B591  (2004)  267.
   %%CITATION = HEP-PH 0403240;%%.

\bibitem{BLM07}
 Victor Braguta, Wolfgang Lucha, and Dmitri Melikhov,
  Phys. Lett. B661  (2008)  354.
   %%CITATION = 0710.5461;%%.

\bibitem{AB08}
 A.~P. Bakulev,
  Phys. Part. Nucl. 40  (2009)  715.
   %%CITATION = 0805.0829;%%.

\bibitem{BMS-APT}
 A.~P. Bakulev, S.~V. Mikhailov, and N.~G. Stefanis,
  Phys. Rev. D72 (2005) 074014, 119908(E);
   %%CITATION = HEP-PH 0506311;%%.
  \textit{ibid.} D75 (2007) 056005;
  \textit{ibid.} D77 (2008) 079901(E).
   %%CITATION = HEP-PH 0607040;%%.

\bibitem{BKS05}
 A.~P. Bakulev, A.~I. Karanikas, and N.~G. Stefanis,
  Phys. Rev. D72  (2005)  074015.
   %%CITATION = HEP-PH/0504275;%%.

\bibitem{FFPI-Cornell}
 C.~J. Bebek \textit{ et~al.},
  Phys. Rev. D9  (1974)  1229;
   %%CITATION = PHRVA,D9,1229;%%.
  \textit{ibid.} D13  (1976)  25;
   %%CITATION = PHRVA,D13,25;%%.
  \textit{ibid.} D17  (1978)  1693.
   %%CITATION = PHRVA,D17,1693;%%.

\bibitem{JLab08II}
 G.~M. Huber \textit{ et~al.},
  Phys. Rev. C78  (2008) 045203.
   %%CITATION = 0809.3052;%%.

\end{thebibliography}
%%\end{document}
%%%%%%%%%%%%%%%%%%%%%%%%%%%%%%%%%%%%%%%%%%%%%%%%%%%%%%%%%%%%%%%%%%%%%%%%%%%%%%%

\end{document}